\documentclass[twocolumn,prl,showpacs,preprintnumbers,amsmath,amssymb]{revtex4}

\usepackage[dvips]{graphicx}
\usepackage{dcolumn}
\usepackage{bm}

\begin{document}

\preprint{}

\title{Fractal Rigidity in Migraine}

\author{Miroslaw Latka}
\email{mirek@if.pwr.wroc.pl}
\homepage{http://www.if.pwr.wroc.pl/~mirek}
\affiliation{
Institute of Physics, Wroclaw University of Technology, Wybrzeze Wyspianskiego 27,
                    50-370 Wroclaw, Poland\\
}%

\author{Marta Glaubic-Latka}
\affiliation{Opole Regional Medical Center
 Al. Witosa 26, 45-401 Opole, Poland 
}\email{mlatka@wcm.opole.pl}
\author{Dariusz Latka}
 \affiliation{Department of Neurosurgery, Opole Regional Medical Center,
              Al. Witosa 26, 45-401 Opole, Poland 
}
 \email{dlatka@wcm.opole.pl}

\author{Bruce J. West}
\email{westb@aro.arl.army.mil}
\affiliation{Mathematics Division, Army Research Office, P.O. Box 12211, Research Triangle, NC 27709-2211, USA
}

\date{January 20, 2003}

\begin{abstract}
We study the middle cerebral artery blood flow velocity (MCAfv) in humans
using transcranial Doppler ultrasonography (TCD).
Scaling properties of time series of the axial flow velocity averaged over a cardiac beat 
interval may be characterized by two exponents. The short time scaling exponent (STSE) determines
the statistical properties of fluctuations of blood flow velocities in short-time intervals
while the Hurst exponent describes the long-term fractal properties.
In many migraineurs the value of the STSE is significantly reduced
and may approach that of the Hurst exponent. This change in dynamical properties
reflects the significant loss of short-term adaptability and 
the overall hyperexcitability of the underlying cerebral blood flow control system.
We call this effect fractal rigidity.

\end{abstract}

\maketitle

Migraine headaches have been the bane of humanity for centuries, afflicting
such notables as Ceasar, Pascal, Kant, Beethoven, Chopin, and Napoleon.
However, their aetiology and pathomechanism have not to date been
satisfactorily elucidated. Herein we demonstrate that the scaling properties of
the time series associated with cerebral blood flow (CBF) significantly  differ
between those of normal healthy individuals
and  migraineurs. The results of our data analysis show that the complex
pathophysiology of migraine frequently leads to hyperexcitability of the cerebral
flow control system.

Physiological signals, such as CBF time series, are typically generated by
complex self-regulatory systems that process inputs with a broad range of
accessible values. Even though this type of time series may fluctuate in an
irregular and complex manner, they frequently exhibit \textit{self-affine}
or \textit{fractal} properties which can be characterized by a single global
parameter -- the fractal dimension $D$ or equivalently the Hurst exponent $H$ 
($D=2-H$)\cite{bass94}.
The salient property of mathematical random fractal process is the existence of
long-range correlations for $H\neq 1/2$. The studies of the cardiac
beat-to-beat variability have shown the existence of strong long-range
correlations in healthy subjects and demonstrated the breakdown of
correlations in disease \cite{peng93} (see also \cite{goldberger99} and
references therein). A similar pattern was observed in fluctuations in the
stride interval in human gait. The strength of correlations was
significantly reduced both by aging and a neurodegenerative disease. This
effect is frequently referred to as the loss of complexity \cite
{hausdorff95,hausdorff97,west98a,walleczek2000}. Complexity decreases with
the convergence of the Hurst exponent on $H=1/2$.

A healthy human brain is perfused by blood flowing laminarly through the
cerebral vessels providing brain tissue with substrates
such as oxygen and glucose. It turns out that CBF is relatively stable with typical values between 45 and
65 ml/100g of brain tissue per second, despite variations in systemic
pressure as large as 100 $Torr$. This phenomenon is known as \textit{%
cerebral autoregulation} and has been thoroughly documented not only in
humans but also in animals \cite{heistad83,paulson90}.
Autoregulation, which is mainly associated with changes in
cerebrovascular resistance (CVR) of small precapillary brain arteries,
is only one of at least four major mechanisms that regulate CBF.  A considerable body of evidence suggests that CBF is
influenced by local cerebral metabolic activity. As metabolic activity increases so does
flow and vice versa. The actual coupling mechanism underlying this \textit{metabolic regulation}
is unknown but most likely it involves certain vasoactive compounds 
(which affect the diameter of cerebral vessels) such as adenosine, potassium, 
prostaglandines which are locally produced in response to metabolic activity.
\textit{External chemical regulation} is predominantly  associated  with the strong influence
of CO$_{2}$ on cerebral vessels. An increase in carbon dioxide arterial content leads to
marked dilation of vessels (vasolidation) which in turn boosts CBF while a decrease produces
mild vasoconstriction and slows down  CBF. The impact of the sympathetic nervous system on CBF
is often ignored but intense sympathetic activity  results  in vasoconstriction.
This type of \textit{neurogenic regulation} can also indirectly affect cerebral flow via
its influence on autoregulation.

The complex cerebral flow regulation mechanisms are supposed to
be influenced or even to be fundamentally altered in many pathological states.
However, despite the significant advances in brain
diagnostic imaging techniques many functional aspects of CBF regulation are
not fully understood \cite{cutrer00}. For example, migraine -- a prevalent, hemicranial
(asymmetric) headache is among the least understood diseases. In \textit{migraine
without aura} attacks may involve
nausea, vomiting, sensitivity to light, sound, or movement. These associated
symptoms distinguish migraine from ordinary tension-type headaches. In about 15\%
of migraineurs headache is preceded by one or more focal neurological symptoms,
collectively known as the aura. In \textit{migraine  with aura}
the associated symptoms may include transient visual disturbances, marching unilateral paresthesias and
numbness or weakness in an extremity or the face, language disturbances, 
and vertigo \cite{ferrari98,goadsby02}. According to the leading hypothesis,
migraine results from a dysfunction of brain-stem or diencephalic nuclei
that normally modulate sensory input and exert neural influence on
\textit{cranial vessels}, see, for example, \cite{goadsby01,goadsby02} and references therein.
Thus, the fundamental question arises as to whether
migraine can significantly influence cerebral hemodynamics.
Some experimental data reveal clear interhemispheric blood flow asymmetry in some parts of
the brain of migraineurs even during headache-free intervals \cite
{battistella90, mirza98, shyhoj89}.

Transcranial Doppler ultrasonography enables high-resolution measurement of
MCA flow velocity. Even though this technique does not allow us to directly
determine CBF values, it may help to elucidate the nature and role of
vascular abnormalities associated with migraine. Some previous studies have
shown significant changes in cerebrovascular reactivity in migraine
patients \cite{heckmann98}. In this work we look for the signature of the
migraine pathology in the scaling properties of the human MCAfv time series.

The dynamical aspects of the cerebral blood flow regulation were recognized by
Zhang \textit{et al.} \cite{zhang98}. Keunen \textit{et al.} \cite{keunen94,
keunen96} applied the attractor reconstruction technique along with the
Grassberger-Procaccia algorithm and the concept of surrogate data to look
for the manifestations of the nonlinear dynamics in continuous waveforms of
TCD signals. Rossitti and Stephensen \cite{rossitti94} used the relative
dispersion of the MCAfv velocity time series to reveal its fractal nature.
West \textit{et al.} \cite{west99a} extended this line of research by taking
into account the more general properties of fractal
time series. Both studies \cite{rossitti94,west99a} showed that the
beat-to-beat variability in the flow velocity has a long-time memory and is
persistent with the average value of the Hurst exponent $H=0.85\pm 0.04$,
a value consistent with that found earlier for interbeat interval 
time series of the human heart. Finally, West \textit{et al} observed
that cerebral blood flow is multifractal in nature \cite{west03}.

\begin{figure}[p]
\includegraphics{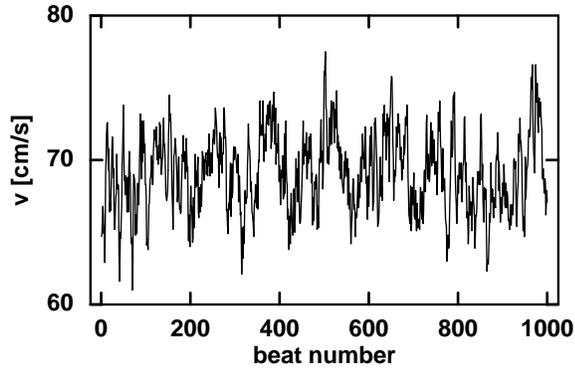}
\caption{MCAfv time series for a healthy subject.}
\label{KarolinaTrace}
\end{figure}

We measured MCAfv using the Multidop T DWL Elektronische Systeme
ultrasonograph. The 2-MHz Doppler probes were placed over the temporal
windows and fixed at a constant angle and position. The measurements were
taken continuously for approximately two hours in the subjects at supine
rest. The study comprised 15 healthy individuals and 33 migraineurs (14
had migraine with aura and the others had migraine without aura).
Migraine was diagnosed according to the guidelines of
the Headache Classification Committee of the International Headache
Society \cite{class}.  An example of a typical measured MCAfv time series is shown
in Fig. \ref{KarolinaTrace} for the first thousand of the recorded beats of the subject's
heart. The total time series has over eight thousand data points for 
a two hour data record.

Successive increments of mathematical fractal random processes are
independent of the time step. They are correlated with the coefficient of
correlation $\rho $ which is determined by the formula $2^{2H}=2+2\rho $.
Thus for $H\neq 1/2$ there exist long-range correlations, that is, $\rho
\neq 0$. It turns out the Hurst exponent also determines the \textit{scaling}
properties of the fractal time series. If $y(t)$ is a fractal process with
Hurst exponent $H,$ then $y_{c}=y(ct)/c^{H}$ is another fractal process with
the same statistics. 
The variance of self-affine time series is proportional to $\Delta t^{2H}$ where
$\Delta t$ is the time interval between measurements. A number of algorithms
which are commonly used to calculate the Hurst exponent are based on this
property.

Herein we employ the detrended fluctuation analysis (DFA) introduced into
the study of biomedical time series by Peng \textit{et al.} \cite{peng94}. 
Let ${\{v_{i}\}}_{i=1}^{N}$ be the experimental time series of the middle
cerebral artery blood flow velocity (MCAfv) $v$. First, the time series is
aggregated: $y(k)=\sum_{i=1}^{k}(v_{i}-\bar{v}),k=1,..,N$, where $\bar{v}$
is the average velocity. Then, for segments of the aggregated time series of
length $n$ the following quantity is calculated: 
\begin{equation}
F(n)=\sqrt{\overline{{{\frac{1}{{n}}}%
\sum_{k=n_{0}}^{n+n_{0}}[y(k)-y_{n_{0}}(k)]^{2}}}},  \label{F}
\end{equation}
where $y_{n_{0}}$ is a least square line fit to the data segment which
starts at $n_{0}$ and ends at $n+n_{0}$. The bar in the above equation
denotes an average over all possible starting points $n_{0}$ of data
segments of length $n$. Thus, for a given data box size $n$, $F(n)$ gives
the characteristic size of fluctuations of the aggregated and detrended time
series. If the aggregated time-series is fractal then $F(n)\sim n^{H},$ so
one obtains the Hurst exponent from a linear least-square fit to $F(n)$ on
double log graph paper. However, West has emphasized \cite{bass94, west99a}
the importance of possible periodic modulations of quantities such as $F(n)$.
These modulations  may be accounted for with the help of the following fit function:
\begin{equation}
F_{X}(n)=n^{H}\exp [\alpha +\lambda \cos (\gamma \ln n)].
\label{fitFunction}
\end{equation}
Here again the Hurst exponent is determined by the slope of the fitting
curve, but now the curve also has a harmonic modulation in the logarithm of
the length $n$ of the data segment.

\begin{figure}[]
\includegraphics{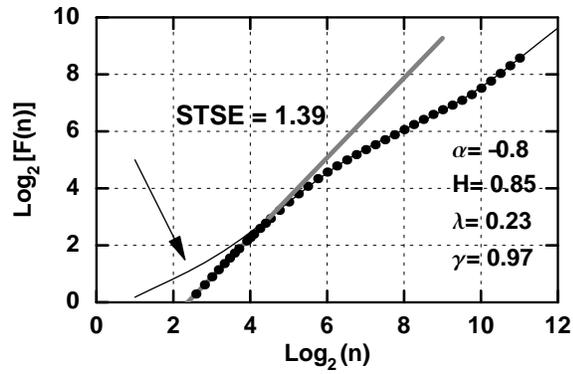}
\caption{Detrended Fluctuation Analysis of the MCAfv time series of a healthy subject.
         The slope of the the grey line gives the value of the short time scaling
         exponent STSE. The parameters of the fit (\ref{fitFunction}) are shown
	 in the right bottom corner of the graph.}
\label{ControlDFA}
\end{figure}

Fig. \ref{ControlDFA} shows the typical DFA analysis for a healthy subject.
The circles in this figure are the calculated values of $F(n)$ and the solid
line is the best renormalization group fit, \textit{cf.} Eqn. (\ref
{fitFunction}). The best-fit parameters are given in the inset.
It is apparent from this graph that fractal properties of the cerebral flow,
as indicated by the quality of the fit, are well pronounced only for
time intervals larger than approximately 32 cardiac beats.
Scaling properties for shorter intervals are distinctly different and may
be characterized by a short time scaling exponent (STSE). The slope of
the grey line in Fig. \ref{ControlDFA} yields the value of the STSE.
The average value of the STSE for the control group was  $1.34 \pm 0.11$
while the average Hurst exponent  was $H=0.80\pm 0.10$. The averaging 
was done over  36 calculations such as the one shown in Fig. \ref{ControlDFA}. We would like
to point out that the values of both exponents are not affected by
a scaling of the time series amplitude and do not depend on the series's mean value
\textit{cf.} (\ref{F}).

\begin{figure}
\includegraphics{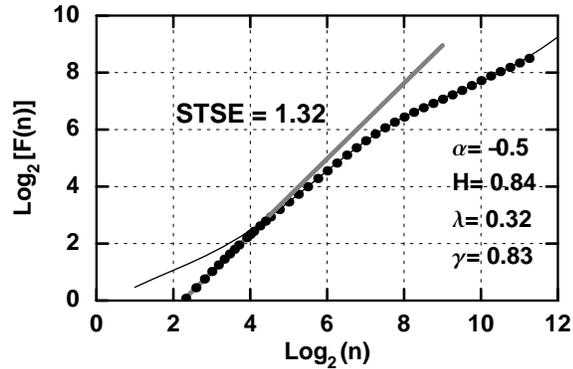}
\caption{DFA of a time series generated by the cerebral blood flow map (CBFM)
         with the following parameters: $b=0.10$, $a=0.05$, $\tau=15$ and $\sigma=0.20$
	 (the time series was scaled to facilitate comparison with the DFA
	 shown in Fig. \ref{ControlDFA}).}
\label{Exper1DFA}
\end{figure}

To elucidate the nature of the two scaling regions characteristic of healthy individuals
let us consider the following one-dimensional map:
\begin{equation}
v_{i+1}=v_{i}-b v_{i} + \frac{a v_{i-\tau}} {1+v_{i-\tau}^{10}} + \xi_{i},
\label{SMGM}
\end{equation}
which models the fluctuations of the blood flow velocity averaged
over a cardiac beat. The map is reminiscent of the Mackey-Glass differential
equation  originally introduced to describe production of
white blood cells \cite{glass88}. The role of the linear term in
the above equation is to dampen out fluctuations. The nonlinear positive
feedback term with delay is the source of long-range correlations. $\xi$
is a random variable which mimics the stochastic component of the cerebral
hemodynamics. In this work we choose $\xi$ to be normally distributed
with zero mean and standard deviation $\sigma$. We call the difference
equation (\ref{SMGM}) a cerebral blood flow map (CBFM).
Fig. \ref{Exper1DFA} shows the DFA of a time series generated by the CBFM
with the parameters $b=0.10$, $a=0.05$, $\tau=15$ and $\sigma=0.20$. The
map clearly exhibits  the characteristics of the experimental data including
the distinct crossover.

\begin{figure}
\includegraphics{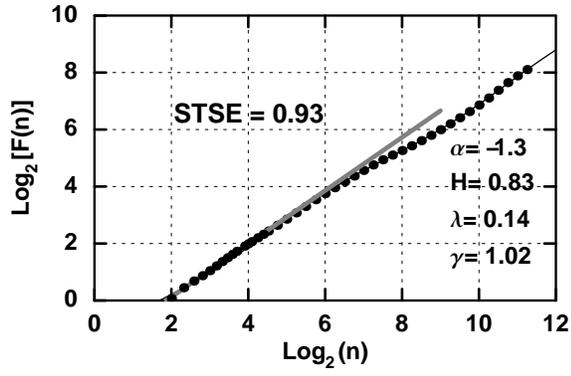}
\caption{DFA of the MCAfv time series of a migraineur without aura. The time
         series was recorded during a headache-free interval.}
\label{NoAuraDFA}
\end{figure}

\begin{figure}
\includegraphics{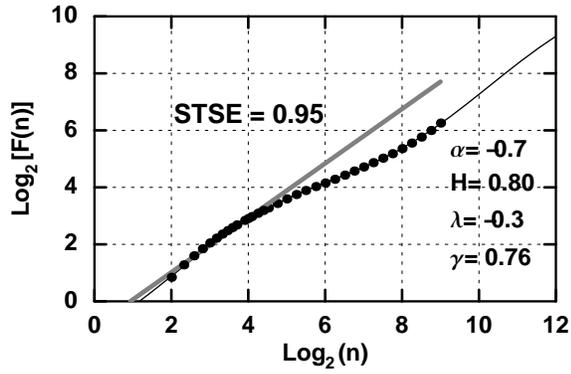}
\caption{DFA of the MCAfv time series of a migraineur with aura.
         The measurement was performed during a headache.}
\label{AuraDFA}
\end{figure}

\begin{figure}
\includegraphics{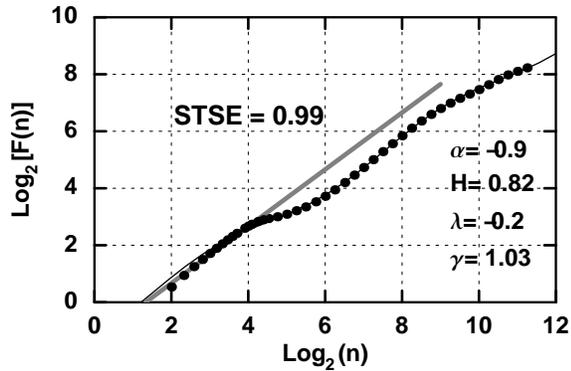}
\caption{DFA of a time series generated by the CBFM with the following
         parameters: $b=0.40$, $a=0.35$, $\tau=15$ and $\sigma=0.20$.}
\label{Rigid1DFA}
\end{figure}

It turns out that the scaling properties of the MCAfv time
series may be profoundly influenced by the migraine pathophysiology.
In our study, for about 40\% of migraineurs the value of the STSE is significantly reduced
and may closely approach the value of the Hurst exponent
\textit{cf.} Figs. \ref{NoAuraDFA}  and \ref{AuraDFA}. On the other hand,
for both the migraineurs with aura and without the average
Hurst exponent is \textit{the same} as that of the control group.
The CBFM provides the insight into dynamical origin of this effect.  The value of the
STSE is dependent mainly on the strength of the damping $b$. With increasing
$b$ the STSE  decreases. However, to maintain approximately constant value
of the Hurst exponent the strength of the positive-feedback
term $a$ must also increase. Fig. \ref{Rigid1DFA} exemplifies this behavior
(the time series  was iterated with the following parameters:
$b=0.45$, $a=0.35$, $\tau=15$ and $\sigma=0.20$). Thus, the reduction of the STSE
observed in the experimental data is the result of the excessive dampening
of the cerebral flow fluctuations and is the manifestation of the significant
loss of adaptability and  overall hyperexcitablity of the underlying regulation system.
We call this novel effect \textit{fractal rigidity}.
We would like to emphasize that hyperexcitabity of the cerebral blood flow control
system seems to be  physiologically consistent with the  reduced activation level of cortical
neurons observed in some  transcranial magnetic stimulation and evoked potential
studies \cite{cutrer00, afra00,ferrari98, goadsby01} and references therein.


\bibliography{FractalRigidity}%

\begin{thebibliography}{27}
\expandafter\ifx\csname natexlab\endcsname\relax\def\natexlab#1{#1}\fi
\expandafter\ifx\csname bibnamefont\endcsname\relax
  \def\bibnamefont#1{#1}\fi
\expandafter\ifx\csname bibfnamefont\endcsname\relax
  \def\bibfnamefont#1{#1}\fi
\expandafter\ifx\csname citenamefont\endcsname\relax
  \def\citenamefont#1{#1}\fi
\expandafter\ifx\csname url\endcsname\relax
  \def\url#1{\texttt{#1}}\fi
\expandafter\ifx\csname urlprefix\endcsname\relax\def\urlprefix{URL }\fi
\providecommand{\bibinfo}[2]{#2}
\providecommand{\eprint}[2][]{\url{#2}}

\bibitem[{\citenamefont{Bassingthwaighte
  et~al.}(1994)\citenamefont{Bassingthwaighte, Liebovitch, and West}}]{bass94}
\bibinfo{author}{\bibfnamefont{J.~B.} \bibnamefont{Bassingthwaighte}},
  \bibinfo{author}{\bibfnamefont{L.~S.} \bibnamefont{Liebovitch}},
  \bibnamefont{and} \bibinfo{author}{\bibfnamefont{B.~J.} \bibnamefont{West}},
  \emph{\bibinfo{title}{Fractal Physiology}} (\bibinfo{publisher}{Oxford
  University Press}, \bibinfo{address}{Oxford}, \bibinfo{year}{1994}).

\bibitem[{\citenamefont{Peng et~al.}(1993)\citenamefont{Peng, Mistus,
  Hausdorff, Havlin, Stanley, and Goldberger}}]{peng93}
\bibinfo{author}{\bibfnamefont{C.~K.} \bibnamefont{Peng}},
  \bibinfo{author}{\bibfnamefont{J.}~\bibnamefont{Mistus}},
  \bibinfo{author}{\bibfnamefont{J.~M.} \bibnamefont{Hausdorff}},
  \bibinfo{author}{\bibfnamefont{S.}~\bibnamefont{Havlin}},
  \bibinfo{author}{\bibfnamefont{H.~E.} \bibnamefont{Stanley}},
  \bibnamefont{and} \bibinfo{author}{\bibfnamefont{A.~L.}
  \bibnamefont{Goldberger}}, \bibinfo{journal}{Phys. Rev. Lett.}
  \textbf{\bibinfo{volume}{70}}, \bibinfo{pages}{1343} (\bibinfo{year}{1993}).

\bibitem[{\citenamefont{Goldberger}(1999)}]{goldberger99}
\bibinfo{author}{\bibfnamefont{A.~L.} \bibnamefont{Goldberger}},
  \emph{\bibinfo{title}{Nonlinear Dynamics, Fractals, and Chaos Theory:
  Implications for Neuroautonomic Heart Rate Control in Health and Disease}}
  (\bibinfo{publisher}{World Health Organization}, \bibinfo{year}{1999}).

\bibitem[{\citenamefont{Hausdorff et~al.}(1995)\citenamefont{Hausdorff, Peng,
  Ladin, Wei, and Goldberger}}]{hausdorff95}
\bibinfo{author}{\bibfnamefont{J.~M.} \bibnamefont{Hausdorff}},
  \bibinfo{author}{\bibfnamefont{C.~K.} \bibnamefont{Peng}},
  \bibinfo{author}{\bibfnamefont{Z.}~\bibnamefont{Ladin}},
  \bibinfo{author}{\bibfnamefont{J.~Y.} \bibnamefont{Wei}}, \bibnamefont{and}
  \bibinfo{author}{\bibfnamefont{A.~L.} \bibnamefont{Goldberger}},
  \bibinfo{journal}{J. Appl. Physiol.} \textbf{\bibinfo{volume}{78}},
  \bibinfo{pages}{349} (\bibinfo{year}{1995}).

\bibitem[{\citenamefont{Hausdorff et~al.}(1997)\citenamefont{Hausdorff,
  Mitchell, Firtion, Peng, Cudkowicz, Wei, and Goldberger}}]{hausdorff97}
\bibinfo{author}{\bibfnamefont{J.~M.} \bibnamefont{Hausdorff}},
  \bibinfo{author}{\bibfnamefont{S.~L.} \bibnamefont{Mitchell}},
  \bibinfo{author}{\bibfnamefont{R.}~\bibnamefont{Firtion}},
  \bibinfo{author}{\bibfnamefont{C.~K.} \bibnamefont{Peng}},
  \bibinfo{author}{\bibfnamefont{M.~E.} \bibnamefont{Cudkowicz}},
  \bibinfo{author}{\bibfnamefont{J.~Y.} \bibnamefont{Wei}}, \bibnamefont{and}
  \bibinfo{author}{\bibfnamefont{A.~L.} \bibnamefont{Goldberger}},
  \bibinfo{journal}{J. Appl. Phys.} pp. \bibinfo{pages}{262--269}
  (\bibinfo{year}{1997}).

\bibitem[{\citenamefont{West and Griffin}(1998)}]{west98a}
\bibinfo{author}{\bibfnamefont{B.~J.} \bibnamefont{West}} \bibnamefont{and}
  \bibinfo{author}{\bibfnamefont{L.}~\bibnamefont{Griffin}},
  \bibinfo{journal}{Fractals} \textbf{\bibinfo{volume}{6}},
  \bibinfo{pages}{101} (\bibinfo{year}{1998}).

\bibitem[{\citenamefont{Peng et~al.}(2000)\citenamefont{Peng, Hausdorff, and
  Goldberger}}]{walleczek2000}
\bibinfo{author}{\bibfnamefont{C.-K.} \bibnamefont{Peng}},
  \bibinfo{author}{\bibfnamefont{J.~M.} \bibnamefont{Hausdorff}},
  \bibnamefont{and} \bibinfo{author}{\bibfnamefont{A.~L.}
  \bibnamefont{Goldberger}}, \emph{\bibinfo{title}{Self-Organized Biological
  Dynamics and Nonlinear Control}} (\bibinfo{publisher}{Cambridge University
  Pres}, \bibinfo{year}{2000}), chap.~\bibinfo{chapter}{3}, pp.
  \bibinfo{pages}{66--96}.

\bibitem[{\citenamefont{Heistad and Kontos}(1983)}]{heistad83}
\bibinfo{author}{\bibfnamefont{D.~D.} \bibnamefont{Heistad}} \bibnamefont{and}
  \bibinfo{author}{\bibfnamefont{H.~A.} \bibnamefont{Kontos}},
  \emph{\bibinfo{title}{Cerebral Circulation}} (\bibinfo{publisher}{Am.
  Physiol. Soc., Bethesda, MD}, \bibinfo{year}{1983}),
  chap.~\bibinfo{chapter}{5}, pp. \bibinfo{pages}{137--182}.

\bibitem[{\citenamefont{Paulson et~al.}(1990)\citenamefont{Paulson,
  Strandgaard, and Edvinsson}}]{paulson90}
\bibinfo{author}{\bibfnamefont{O.~B.} \bibnamefont{Paulson}},
  \bibinfo{author}{\bibfnamefont{S.}~\bibnamefont{Strandgaard}},
  \bibnamefont{and}
  \bibinfo{author}{\bibfnamefont{L.}~\bibnamefont{Edvinsson}},
  \bibinfo{journal}{Cerebrovasc. Brain Metab. Res.}
  \textbf{\bibinfo{volume}{2}}, \bibinfo{pages}{161} (\bibinfo{year}{1990}).

\bibitem[{\citenamefont{Cutrer et~al.}(2000)\citenamefont{Cutrer, O'Donnell,
  and del Rio}}]{cutrer00}
\bibinfo{author}{\bibfnamefont{F.~M.} \bibnamefont{Cutrer}},
  \bibinfo{author}{\bibfnamefont{A.}~\bibnamefont{O'Donnell}},
  \bibnamefont{and} \bibinfo{author}{\bibfnamefont{M.~S.~S.} \bibnamefont{del
  Rio}}, \bibinfo{journal}{Neurology} \textbf{\bibinfo{volume}{55}},
  \bibinfo{pages}{S36} (\bibinfo{year}{2000}).

\bibitem[{\citenamefont{Ferrari}(1998)}]{ferrari98}
\bibinfo{author}{\bibfnamefont{M.~D.} \bibnamefont{Ferrari}},
  \bibinfo{journal}{Lancet} \textbf{\bibinfo{volume}{351}},
  \bibinfo{pages}{1043} (\bibinfo{year}{1998}).

\bibitem[{\citenamefont{Goadsby et~al.}(2002)\citenamefont{Goadsby, Lipton, and
  Ferrari}}]{goadsby02}
\bibinfo{author}{\bibfnamefont{P.~J.} \bibnamefont{Goadsby}},
  \bibinfo{author}{\bibfnamefont{R.~B.} \bibnamefont{Lipton}},
  \bibnamefont{and} \bibinfo{author}{\bibfnamefont{M.~D.}
  \bibnamefont{Ferrari}}, \bibinfo{journal}{N. Engl J Med}
  \textbf{\bibinfo{volume}{346}}, \bibinfo{pages}{257} (\bibinfo{year}{2002}).

\bibitem[{\citenamefont{Goadsby}(2001)}]{goadsby01}
\bibinfo{author}{\bibfnamefont{P.~J.} \bibnamefont{Goadsby}},
  \emph{\bibinfo{title}{Wolff's headache and other head pain}}
  (\bibinfo{publisher}{Oxford University Press}, \bibinfo{year}{2001}), chap.
  \bibinfo{chapter}{Pathophysiology of headache}, pp. \bibinfo{pages}{57--72}.

\bibitem[{\citenamefont{Battistella}(1990)}]{battistella90}
\bibinfo{author}{\bibfnamefont{P.~A.} \bibnamefont{Battistella}},
  \bibinfo{journal}{Headache} \textbf{\bibinfo{volume}{30}},
  \bibinfo{pages}{646} (\bibinfo{year}{1990}).

\bibitem[{\citenamefont{Mirza}(1998)}]{mirza98}
\bibinfo{author}{\bibfnamefont{M.}~\bibnamefont{Mirza}}, \bibinfo{journal}{Acta
  Neurol Belg} \textbf{\bibinfo{volume}{98}}, \bibinfo{pages}{190}
  (\bibinfo{year}{1998}).

\bibitem[{\citenamefont{Olsen and Lassen}(1989)}]{shyhoj89}
\bibinfo{author}{\bibfnamefont{T.~S.} \bibnamefont{Olsen}} \bibnamefont{and}
  \bibinfo{author}{\bibfnamefont{N.}~\bibnamefont{Lassen}},
  \bibinfo{journal}{Headache} \textbf{\bibinfo{volume}{31}},
  \bibinfo{pages}{49} (\bibinfo{year}{1989}).

\bibitem[{\citenamefont{Heckmann}(1998)}]{heckmann98}
\bibinfo{author}{\bibfnamefont{J.~G.} \bibnamefont{Heckmann}},
  \bibinfo{journal}{Cephalalgia} \textbf{\bibinfo{volume}{18}},
  \bibinfo{pages}{133} (\bibinfo{year}{1998}).

\bibitem[{\citenamefont{Zhang et~al.}(1999)\citenamefont{Zhang, Zuckerman,
  Giller, and Levine}}]{zhang98}
\bibinfo{author}{\bibfnamefont{R.}~\bibnamefont{Zhang}},
  \bibinfo{author}{\bibfnamefont{J.~H.} \bibnamefont{Zuckerman}},
  \bibinfo{author}{\bibfnamefont{C.}~\bibnamefont{Giller}}, \bibnamefont{and}
  \bibinfo{author}{\bibfnamefont{B.~D.} \bibnamefont{Levine}},
  \bibinfo{journal}{Am. J. Physiol.} \textbf{\bibinfo{volume}{274}},
  \bibinfo{pages}{H233} (\bibinfo{year}{1999}).

\bibitem[{\citenamefont{Keunen et~al.}(1994)\citenamefont{Keunen, Pijlman,
  Visee, Vliegen, Tavy, and Stam}}]{keunen94}
\bibinfo{author}{\bibfnamefont{R.~W.~M.} \bibnamefont{Keunen}},
  \bibinfo{author}{\bibfnamefont{H.}~\bibnamefont{Pijlman}},
  \bibinfo{author}{\bibfnamefont{H.~F.} \bibnamefont{Visee}},
  \bibinfo{author}{\bibfnamefont{J.~H.~R.} \bibnamefont{Vliegen}},
  \bibinfo{author}{\bibfnamefont{D.~L.~T.} \bibnamefont{Tavy}},
  \bibnamefont{and} \bibinfo{author}{\bibfnamefont{C.~J.} \bibnamefont{Stam}},
  \bibinfo{journal}{Neurophys. Res.} \textbf{\bibinfo{volume}{16}},
  \bibinfo{pages}{353} (\bibinfo{year}{1994}).

\bibitem[{\citenamefont{Keunen et~al.}(1996)\citenamefont{Keunen, Vliegen,
  Stam, and Tavy}}]{keunen96}
\bibinfo{author}{\bibfnamefont{R.~W.~M.} \bibnamefont{Keunen}},
  \bibinfo{author}{\bibfnamefont{J.~H.~R.} \bibnamefont{Vliegen}},
  \bibinfo{author}{\bibfnamefont{C.~J.} \bibnamefont{Stam}}, \bibnamefont{and}
  \bibinfo{author}{\bibfnamefont{D.~L.~T.} \bibnamefont{Tavy}},
  \bibinfo{journal}{Ultrasound Med. Biol.} \textbf{\bibinfo{volume}{22}},
  \bibinfo{pages}{353} (\bibinfo{year}{1996}).

\bibitem[{\citenamefont{Rossitti and Stephensen}(1994)}]{rossitti94}
\bibinfo{author}{\bibfnamefont{S.}~\bibnamefont{Rossitti}} \bibnamefont{and}
  \bibinfo{author}{\bibfnamefont{H.}~\bibnamefont{Stephensen}},
  \bibinfo{journal}{Acta Physio. Scand.} \textbf{\bibinfo{volume}{151}},
  \bibinfo{pages}{191} (\bibinfo{year}{1994}).

\bibitem[{\citenamefont{West et~al.}(1999)\citenamefont{West, Zhang, Sanders,
  Zuckerman, and Levine}}]{west99a}
\bibinfo{author}{\bibfnamefont{B.~J.} \bibnamefont{West}},
  \bibinfo{author}{\bibfnamefont{R.}~\bibnamefont{Zhang}},
  \bibinfo{author}{\bibfnamefont{A.~W.} \bibnamefont{Sanders}},
  \bibinfo{author}{\bibfnamefont{J.~H.} \bibnamefont{Zuckerman}},
  \bibnamefont{and} \bibinfo{author}{\bibfnamefont{B.~D.}
  \bibnamefont{Levine}}, \bibinfo{journal}{Phys. Rev. E}
  \textbf{\bibinfo{volume}{59}}, \bibinfo{pages}{3492} (\bibinfo{year}{1999}).

\bibitem[{\citenamefont{West et~al.}(2003)\citenamefont{West, Latka,
  Glaubic-Latka, and Latka}}]{west03}
\bibinfo{author}{\bibfnamefont{B.~J.} \bibnamefont{West}},
  \bibinfo{author}{\bibfnamefont{M.}~\bibnamefont{Latka}},
  \bibinfo{author}{\bibfnamefont{M.}~\bibnamefont{Glaubic-Latka}},
  \bibnamefont{and} \bibinfo{author}{\bibfnamefont{D.}~\bibnamefont{Latka}},
  \bibinfo{journal}{Physica A} \textbf{\bibinfo{volume}{318}},
  \bibinfo{pages}{431} (\bibinfo{year}{2003}).

\bibitem[{\citenamefont{of~the International Headache~Society}(1988)}]{class}
\bibinfo{author}{\bibfnamefont{H.~C.~C.} \bibnamefont{of~the International
  Headache~Society}}, \bibinfo{journal}{Cephalalgia}
  \textbf{\bibinfo{volume}{8}}, \bibinfo{pages}{10} (\bibinfo{year}{1988}).

\bibitem[{\citenamefont{Peng et~al.}(1994)\citenamefont{Peng, Buldyrev, Havlin,
  Simons, Stanley, and Goldberger}}]{peng94}
\bibinfo{author}{\bibfnamefont{C.~K.} \bibnamefont{Peng}},
  \bibinfo{author}{\bibfnamefont{S.~V.} \bibnamefont{Buldyrev}},
  \bibinfo{author}{\bibfnamefont{S.}~\bibnamefont{Havlin}},
  \bibinfo{author}{\bibfnamefont{M.}~\bibnamefont{Simons}},
  \bibinfo{author}{\bibfnamefont{H.~E.} \bibnamefont{Stanley}},
  \bibnamefont{and} \bibinfo{author}{\bibfnamefont{A.~L.}
  \bibnamefont{Goldberger}}, \bibinfo{journal}{Phys. Rev. E}
  \textbf{\bibinfo{volume}{49}}, \bibinfo{pages}{1685} (\bibinfo{year}{1994}).

\bibitem[{\citenamefont{Glass and Mackey}(1988)}]{glass88}
\bibinfo{author}{\bibfnamefont{L.}~\bibnamefont{Glass}} \bibnamefont{and}
  \bibinfo{author}{\bibfnamefont{M.~C.} \bibnamefont{Mackey}},
  \emph{\bibinfo{title}{From Clocks to Chaos, The Rhythms of Life}}
  (\bibinfo{publisher}{Princeton University Pres, NJ}, \bibinfo{year}{1988}).

\bibitem[{\citenamefont{Afra et~al.}(2000)\citenamefont{Afra, Cecchini, Sandor,
  and Schoenen}}]{afra00}
\bibinfo{author}{\bibfnamefont{J.}~\bibnamefont{Afra}},
  \bibinfo{author}{\bibfnamefont{A.~P.} \bibnamefont{Cecchini}},
  \bibinfo{author}{\bibfnamefont{P.~S.} \bibnamefont{Sandor}},
  \bibnamefont{and} \bibinfo{author}{\bibfnamefont{J.}~\bibnamefont{Schoenen}},
  \bibinfo{journal}{Clinical Neurophysiology} \textbf{\bibinfo{volume}{111}},
  \bibinfo{pages}{1124} (\bibinfo{year}{2000}).

\end{thebibliography}

\end{document}